\documentclass[runningheads]{llncs}
\usepackage[english]{babel}
\usepackage[T1]{fontenc}
\usepackage[utf8]{inputenc}
\DeclareUnicodeCharacter{00AE}{\textsuperscript{\textregistered}}
\usepackage{graphicx}
\usepackage{amsmath}
\usepackage{amssymb}
\usepackage{url}
\usepackage{booktabs}
\usepackage{tabularx}
\usepackage{array}
\usepackage{soul}
\usepackage{cite}
\usepackage{multirow}
\usepackage{makecell}
\usepackage{microtype}
\usepackage{placeins}
\usepackage{xcolor}
\newcolumntype{L}[1]{>{\raggedright\arraybackslash}p{#1}}
\newcolumntype{Y}{>{\raggedright\arraybackslash}X}

\IfFileExists{generated/numbers.tex}{% Generated by experiment/generate.py. Do not edit by hand.
\newif\ifExperimentResultsReady
\ExperimentResultsReadytrue

\newcommand{\ModelOneName}{\textsf{gpt-4o-mini}}
\newcommand{\ModelTwoName}{\textsf{gpt-4.1}}

\newcommand{\ModelCount}{2}
\newcommand{\RunsPerModel}{40}
\newcommand{\TotalLiveRuns}{80}

\newcommand{\PrimaryConditionName}{Python null-pointer-dereference toy benchmark}
\newcommand{\BaselineRootState}{Pass with Exceptions}
\newcommand{\StrictRootState}{Fail}
\newcommand{\ModelTwoRootState}{Inconclusive}
\newcommand{\ReplayStatus}{reproduced}

\newcommand{\MaxTestedContexts}{1{,}000}

\newcommand{\GateBudget}{5\,s}

}{%
  \newif\ifExperimentResultsReady
  \ExperimentResultsReadyfalse
}
\providecommand{\ModelOneName}{\texttt{gpt-4o-mini}}
\providecommand{\ModelTwoName}{\texttt{gpt-4.1}}
\providecommand{\ModelCount}{2}
\providecommand{\RunsPerModel}{40}
\providecommand{\TotalLiveRuns}{80}
\providecommand{\PrimaryConditionName}{Python null-pointer-dereference toy benchmark}
\providecommand{\BaselineRootState}{\textsf{TBD}}
\providecommand{\StrictRootState}{\textsf{TBD}}
\providecommand{\ModelTwoRootState}{\textsf{TBD}}
\providecommand{\ReplayStatus}{\textsf{TBD}}

\def\PolicyLatencyMax{1.1\,ms}
\def\ScaleLatencyMax{1.62\,s}
\def\MaxTestedContexts{1,000}
\def\GateBudget{5\,s}

\AtBeginDocument{%
  \ifExperimentResultsReady\else
    \PackageWarningNoLine{TAIP paper}{Generated experiment metadata are not marked ready. Reconcile the reporting snapshot with the original project before submission}%
  \fi
}

\newcommand{\figureplaceholder}[2]{%
  \fbox{\parbox[c][#1][c]{0.94\linewidth}{\centering\small #2}}%
}

\begin{document}

\title{Continuous Assurance of Agentic Security Auditors for Software Delivery Decision Gates}
\titlerunning{Continuous Assurance for Agentic Security Auditors}

\author{
Guy Lupo\inst{1}\thanks{Corresponding author: glupo@swin.edu.au}
\and Nguyen Hung Nguyen\inst{1}
\and Viet Vo\inst{1}
\and M.A.P. Chamikara\inst{3}
\and Guangdong Bai\inst{2}
\and Nazatul Haque Sultan\inst{3}
\and Alsharif Abuadbba\inst{3}
}

\authorrunning{G. Lupo et al.}

\institute{
Swinburne University of Technology, Melbourne, Australia\\
\email{\{glupo,nguyenhungnguyen,vvo\}@swin.edu.au}
\and
City University of Hong Kong, Kowloon Tong, Hong Kong\\
\email{g.bai@cityu.edu.hk}
\and
CSIRO, Australia\\
\email{\{chamikara.arachchige,nazatul.sultan,sharif.abuadbba\}@csiro.au}
}

\maketitle

\begin{abstract}
Large language model (LLM)-based repository auditors are increasingly deployed as security controls within continuous integration (CI) pipelines, where their findings admit, block, or delay software changes. As Agentic Software Development Life Cycle (SDLC) Security Controls, their non-deterministic behaviour changes the evidence, while organisational risk appetite and jurisdictional or data-sovereignty policy change its interpretation. Point-in-time audits therefore cannot maintain current assurance for merge decisions. Existing approaches do not fully address this complexity: life cycle audits are episodic, runtime operations focus on individual operations, and provenance systems record evidence without determining whether it remains admissible under the active configuration and policy. Consequently, existing approaches cannot provide a context-bound, continuously recomputed assurance verdict within the time budget of a software-delivery gate. 

We propose the Policy--Evidence--Execution Separation Pattern, implemented by the executable Trustworthy AI Posture (TAIP) Assurance Engine and operated as Continuous Control Posture Assurance (CCPA). By separating policy from stable execution and binding admitted evidence to a versioned Posture Tree, the same assurance logic operates unchanged across models, environments and policy profiles, allowing assurance to remain current as AI-enabled security controls evolve.

The approach is evaluated using the unmodified RepoAudit repository auditor on a fixed Python Null Pointer Dereference benchmark. A retained evidence repository comprising \textbf{80} RepoAudit executions across two OpenAI model configurations (\texttt{gpt-4o-mini} and \texttt{gpt-4.1}) is first established. TAIP then measures the time required to recompute and publish an updated assurance posture following policy, evidence and model-context changes before evaluating the same assurance computation across increasing numbers of independent Decision Gateway contexts.

The maximum observed policy-to-posture latency was \textbf{\PolicyLatencyMax{}} (rounded) across three policy-class cycles in one execution. At \textbf{\MaxTestedContexts{}} independent assurance contexts, full policy-triggered recomputation with one worker recorded a maximum aggregate refresh of \textbf{\ScaleLatencyMax{}}, below the predeclared \textbf{\GateBudget{}} Decision Gateway budget. These single-host measurements concern assurance over retained evidence and exclude RepoAudit execution and provider inference.

\end{abstract}

\section{Introduction}
\label{sec:introduction}

LLM-based repository auditors are moving into continuous integration and release workflows. They inspect repositories, choose code paths, invoke analysis tools, validate candidate defects, and emit findings that may influence merge decisions. When such findings can admit or block code, the auditor is no longer merely an audit tool; it functions as an Agentic Software Development Life Cycle Security Control. RepoAudit \cite{guo_repoaudit_2025} is a concrete example: it performs repository-level bug discovery via model-guided exploration, memory, data-flow reasoning, and path validation. Although its assigned function is detective, its behaviour is model-dependent and partially non-deterministic \cite{national_institute_of_standards_and_technology_nist_2024}.

Traditional security assurance assumes that a control's behaviour remains sufficiently stable for an earlier assessment to remain valid. This assumption generally holds for deterministic static analysers, whose implementation defines a predictable relationship between inputs and outputs. Their assurance status can therefore be retained until the binary, ruleset, scope, or operating environment changes. It does not hold as reliably for LLM-backed controls: a provider can swap the underlying model behind an unchanged name, alter routing, impose rate limits, or suffer transient degradation while the control continues to return findings. The delivery pipeline may therefore observe an apparently successful run even when the configuration that justified adoption is not the configuration that executed. These conditions create two independent paths of change. Technical events can alter the evidence produced (e.g., via the requested model, provider routing, repository state, prompt, validator, runtime conditions, or evidence window). Policy events can change the interpretation of unchanged evidence (e.g., via risk appetite, advisory versus blocking use, provider restrictions, retention duties, or jurisdictional and data-sovereignty requirements). Findings, logs, token records, timing, and integrity manifests make the control observable, but observability alone does not establish that it remains adequate, effective, and sustainable for the active gate. Evaluation becomes assurance evidence only when it is bound to a scoped claim, configuration, evidence window, and intended decision \cite{staufer_audit_2025,herrera-poyatos_responsible_2025,haugen_assurance_2025}.

Existing work addresses individual parts of the assurance chain, but no single approach provides continuous, context-bound assurance for SDLC decision gates.
Frameworks such as SMACTR, Ethics-Based Auditing, capAI, GAFAI, and Z-Inspection organise the assessment of AI systems around defined scope, roles, artefacts, tests, and reporting procedures \cite{raji_closing_2020,mokander_ethics-based_2021,floridi_capai_2022,markert_gafai_2022,zicari_z-inspection_2021}.
Their operative unit is an audit engagement or lifecycle review whose interpretation remains dependent on human judgement. AgentSentinel and SmartAuditFlow can intercept an operation and apply a configured security audit before enforcement \cite{hu_agentsentinel_2025}, functioning as an automated detective assurance with a feedback loop.  RepoAudit, RFCAudit, and SpecAuditor automate target-specific audit work and produce findings or validation evidence focused on operational controls \cite{guo_repoaudit_2025,zheng_rfcaudit_2025,wei_smartauditflow_2025,lin_specauditor_2026-1}.
These systems materially improve testing and local control action, but their published designs do not bind longitudinal evidence about the auditor's (i) adequacy (is it the right tool for the job), (ii) detection efficacy (is it working, available, and maintaining integrity), and (iii) cost (is it affordable over time) to a current organisational claim in a way that supports a decision gate consequence. Provenance and bill-of-materials mechanisms preserve evidence for replay and investigation, but they do not select the active claim, policy, or decision budget \cite{souza_prov-agent_2025,naja_using_2022,radanliev_operationalising_2026}.
This gap becomes a scaling constraint in agentic SDLCs: manual review is limited by reviewer capacity, while decision contexts increase with repositories, pipelines, and autonomous coding workstreams. Each gate still requires a current verdict, the governing policy version, admissible evidence for the active configuration, and publication within its decision budget. Without these properties, it either reuses stale assurance or applies an unexamined default. Existing continuous auditing and posture-management approaches provide temporal monitoring, but not the context-bound integration needed to support changing agentic controls at software delivery gates \cite{minkkinen_continuous_2022,dempsey_information_2011,al-karaki_gosafe_2022}. Continuous assurance instead allows a gate to wait for a current posture. If the wait budget expires, the result is \textsc{Inconclusive}, and policy determines whether to block the change or permit it through a recorded exception. Timely posture computation is therefore essential for machine-speed SDLC decisions \cite{sinan_integrating_2025,minkkinen_continuous_2022,dempsey_information_2011}.

We address this gap with the \emph{Policy--Evidence--Execution Separation Pattern}, which separates three concerns that evolve independently. Policy defines the conditions under which a control may be relied upon and the consequence of each posture state; evidence records observations, provenance, and configuration; and execution applies stable assurance semantics. The pattern represents an assurance claim as a versioned Posture Tree whose leaves bind admissible evidence and whose internal nodes aggregate results through a uniform Composite interface. Consequently, models, operating contexts, and policy profiles can change without requiring corresponding changes to the assurance engine. We realise the pattern in the \emph{Trustworthy AI Posture} (TAIP) Assurance Engine and operate it through \emph{Continuous Control Posture Assurance} (CCPA). TAIP is placed around unmodified RepoAudit, which serves as a representative agentic SDLC security control. The assurance harness responds to changes in admitted evidence, model context, and policy profile by recomputing the resulting control posture. We use \emph{time-to-posture} to denote the elapsed time from event admission to publication of the corresponding Posture Card at the local gate interface.

This study addresses the following research questions:

\begin{description}
    \item[\textbf{RQ1.}] What software architecture enables current and scalable assurance of agentic SDLC security controls at software delivery decision gates?

    \item[\textbf{RQ2.}] Can CCPA recompute policy-bound control posture within the decision budget of a software delivery gate, and how does its performance scale with evidence depth and the number of independent gate contexts?
\end{description}

\begin{figure}[t]
\centering
\IfFileExists{figures/figure1-assurance-context.png}{%
  \includegraphics[width=\linewidth]{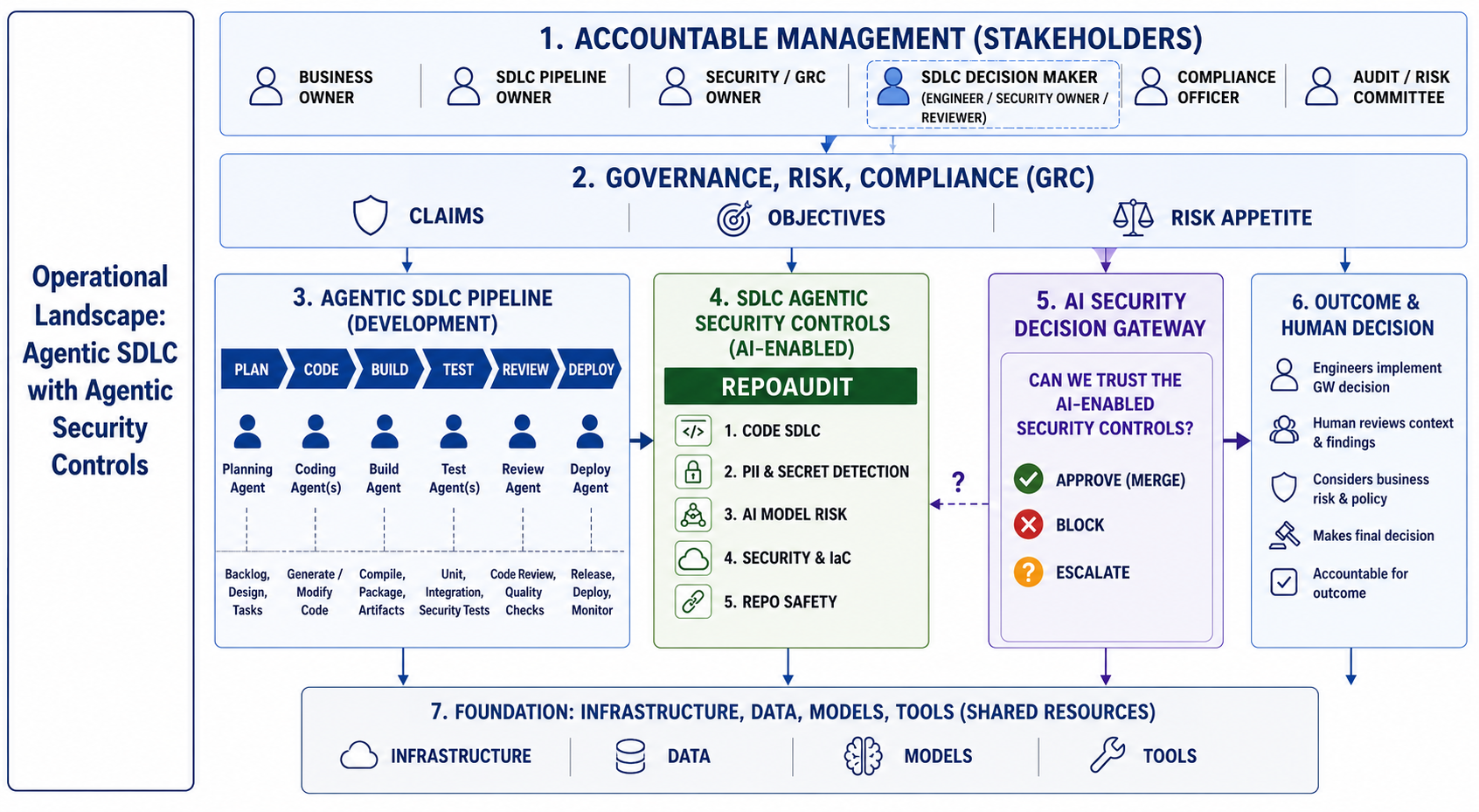}%
}{%
  \figureplaceholder{58mm}{\textbf{Figure 1 scaffold: assurance context.} Operational SDLC contexts feed RepoAudit. Manual audits, runtime checks, and TAIP occupy different assurance roles. TAIP binds the selected policy to evidence from the active control context and publishes one current gate decision.}%
}
\caption{Assurance context used throughout the paper. Manual lifecycle audit and automated checking produce point-in-time or local evidence. TAIP occupies the assurance layer, binding the versioned claim and policy profile to control-test evidence before publishing a current posture for the SDLC decision gate.}
\label{fig:assurance_context}
\end{figure}
RQ2 evaluates one property that is necessary for continuous assurance at an
SDLC Decision Gateway: the time required to replace an inherited assurance
decision with a current posture after a relevant change. The policy-change
experiment provides the primary measurement because it isolates the TAIP
assurance computation from RepoAudit execution, model inference and evidence
collection. Additional experiments demonstrate that the same computation
admits newly collected evidence, invalidates evidence generated under
\ModelOneName{} when the operating context changes to
\ModelTwoName{}, and republishes a current posture without re-executing the
underlying security control.

\paragraph{Answer to RQ2.}

Within the tested single-host environment, the maximum observed policy-to-posture latency was \textbf{\PolicyLatencyMax{}} (rounded) for one Context. At \textbf{\MaxTestedContexts{}} independent assurance contexts, full policy-triggered recomputation with one worker recorded a maximum aggregate refresh of \textbf{\ScaleLatencyMax{}}. Both values were below the predeclared \textbf{\GateBudget{}} Decision Gateway budget, supporting temporal adequacy within the tested evidence and execution conditions.

The experiment is intentionally bounded. It evaluates the responsiveness of
the assurance mechanism rather than RepoAudit detection accuracy,
distributed execution, comparative performance against alternative assurance
approaches, or production-scale deployment. Those questions remain the
subject of future work.

\subsection{Contributions}
The paper makes three primary contributions. 

\begin{enumerate}
    \item \textbf{A separation pattern for assuring agentic SDLC security controls.}
    We introduce the \emph{Policy--Evidence--Execution Separation Pattern}, which separates organisational policy, control evidence, and assurance execution. The pattern uses versioned Posture Trees, context-bound evidence admission, mandatory assurance branches, and exception-preserving aggregation.

    \item \textbf{An executable implementation.}
    We implement the pattern in the \emph{TAIP Assurance Engine} and integrate it with unmodified RepoAudit through a \emph{Continuous Control Posture Assurance} (CCPA) event loop. The implementation supports evidence admission, policy re-evaluation, context invalidation, posture publication, and replay.
    \item \textbf{A reproducible decision-gate evaluation.}
    We evaluate assurance over retained evidence, reporting maximum observed latency of \textbf{\PolicyLatencyMax{}} (rounded) for one Context and \textbf{\ScaleLatencyMax{}} for aggregate refresh across \textbf{\MaxTestedContexts{}} contexts with one worker. These measurements isolate assurance computation from live RepoAudit execution.
\end{enumerate}

\section{Related Work and Assurance-Layer Gap}
\label{sec:related}

Figure~\ref{fig:assurance_context} separates the operational SDLC, the detective control, control testing and evidence production, and the assurance decision. The comparison below focuses on the final layer: whether prior work (i) binds evidence to the active context and policy, (ii) aggregates that evidence into a single current claim decision, (iii) represents absence of a defensible decision, and (iv) returns a result within an SDLC gate budget.

\subsection{Lifecycle audit and assurance processes}

SMACTR and related internal-audit frameworks structure scoping, artefact collection, testing, reflection, and reporting; Ethics-Based Auditing, capAI, GAFAI, and Z-Inspection translate principles or regulatory expectations into review procedures \cite{raji_closing_2020,mokander_ethics-based_2021,floridi_capai_2022,markert_gafai_2022,zicari_z-inspection_2021}. They remain well suited to adoption review, periodic reassessment, accountability, and sign-off. Their primary outputs are engagement reports, checklists, scorecards, or assurance arguments produced through expert judgement. However, rerunning an engagement after each model, evidence, or policy event does not yield a bounded event-to-verdict path for every affected merge gate. Requirements-engineering work similarly cautions that checklist completion can obscure context-dependent and evolving trustworthiness \cite{donati_beyond_2025}.

Evidence-based and dynamic assurance-case research strengthens the link between claims, arguments, and evidence \cite{haugen_assurance_2025,sabuncuoglu_justified_2025}. It emphasises that evidence is meaningful only relative to the claim it supports and that assurance must evolve with the system. The remaining software-engineering question is how to instantiate such arguments for a population of SDLC gates, recompute them after discrete events, and publish a verdict with measured event-to-publication latency.

\subsection{Agentic audit tools and runtime controls}

RepoAudit, RFCAudit, SmartAuditFlow, and SpecAuditor automate different audit targets and reasoning strategies \cite{guo_repoaudit_2025,zheng_rfcaudit_2025,wei_smartauditflow_2025,lin_specauditor_2026-1}. RepoAudit explores repository paths and validates candidate defects; RFCAudit checks implementations against normative RFC requirements; SmartAuditFlow adapts plan-and-execute reasoning for smart-contract auditing; and SpecAuditor derives and applies audit specifications. These systems produce findings and target-specific validation evidence. From the assurance-layer perspective, however, they typically stop before evaluating a versioned organisational claim over multiple evidence families under the currently active policy. They function as active testers, whose outputs must still be interpreted by a mechanism that decides whether the auditor itself remains fit to influence a merge gate.

AgentSentinel operates at a different layer. It traces computer-use agent operations, suspends a sensitive operation, and runs a context-aware security audit before enforcement \cite{hu_agentsentinel_2025}. This demonstrates low-latency interception and local blocking based on whether an operation violates configured security conditions. It does not, however, aggregate longitudinal evidence to determine whether the monitor or auditor remains adequate, effective, and sustainable under the active model, evidence window, and policy profile. Its local allow or block is therefore a control action rather than a posture decision over the control.

\subsection{Evidence infrastructure and continuous posture precedents}

Provenance graphs, semantic accountability models, and AI bill-of-materials proposals improve traceability across prompts, actions, actors, components, and evidence \cite{souza_prov-agent_2025,naja_using_2022,radanliev_operationalising_2026}. These records support forensics, replay, and auditability. They remain policy-neutral until another mechanism selects the claim, checks scope and freshness, applies thresholds and mandatory rules, preserves exceptions, and publishes a gate consequence. Agentic auditability proposals reinforce integrity, coverage, temporal coherence, verifiability, and governance alignment at this evidence boundary \cite{phiri_creating_2025}.

NIST AI RMF supplies the Test--Evaluate--Verify--Validate vocabulary used in TAIP, while control frameworks supply the continuing adequacy, operating effectiveness, and viability criteria \cite{tabassi_ai_2023,calagna_applying_2021}. Information-security continuous monitoring and posture management demonstrate the value of recomputing control status as infrastructure changes \cite{dempsey_information_2011,al-karaki_gosafe_2022,minkkinen_continuous_2022}. They provide the temporal precedent, but they do not define a context-bound assurance object for a non-deterministic SDLC auditor, a first-class \textsc{Inconclusive} state, or a policy-specific gate action evaluated against a declared waiting budget.

\begin{table}[t]
\caption{Related work positioned by the artefact delivered to the assurance layer.}
\label{tab:related_assurance_layer}
\centering
\small
\begin{tabularx}{\linewidth}{L{0.39\linewidth}Y}
\toprule
Work class, examples, and primary output & Missing assurance-layer function at an SDLC gate \\
\midrule
Lifecycle audit: SMACTR, capAI, GAFAI, Z-Inspection. Report, checklist, scorecard, or sign-off. & No bounded recomputation after each policy, model, evidence, or expiry event. \\
Dynamic assurance cases. Versioned claim, argument, and supporting evidence. & No measured event loop that publishes one current result across a population of merge gates. \\
Agentic audit: RepoAudit, RFCAudit, SmartAuditFlow, SpecAuditor. Findings and target-specific validation evidence. & No claim-level aggregation about the audit control's own continuing fitness. \\
Runtime control: AgentSentinel. Local allow, suspend, or block for an intercepted operation. & No longitudinal multi-objective posture over the control under a versioned policy. \\
Provenance and AIBOM infrastructure. Traces, manifests, component and actor records. & No active context admission, root state, gate consequence, or decision budget. \\
Continuous monitoring and posture management. Recurring indicators and control status. & No context-bound assurance object and explicit uncertainty consequence for an agentic SDLC control. \\
\bottomrule
\end{tabularx}
\end{table}

\subsection{Assurance-layer integration gap}

The gap is not another defect detector. It is a software layer that turns heterogeneous control-test evidence into one accountable and current decision. That layer must separate fact from judgement, bind each observation to the scope and configuration that produced it, reject incompatible or stale evidence before aggregation, retain failed and inconclusive branches, and externalise the consequence of uncertainty in policy. It must also publish inside the gate's decision budget. TAIP does not replace the work in Table~\ref{tab:related_assurance_layer}; it converts its evidence into a forward-looking control posture that can govern an SDLC decision.

\section{RQ1: The Policy--Evidence--Execution Separation Pattern}
\label{sec:rq1}

\subsection{Pattern Structure: Composite Posture and Recursive TEVV}

RQ1 asks how assurance can remain current while an agentic control,
its evidence, and its policy change independently. The
\emph{Policy--Evidence--Execution Separation Pattern} assigns each
concern a boundary: policy defines \emph{what must hold}, evidence
records \emph{what occurred}, and execution defines \emph{how a
claim decision is computed}
\cite{lupo_trustworthy_2026,agarwal_compliance-as-code_2022}.

\paragraph{Posture Tree and Composite objects.}
A posture instance is $P=(T,\pi,\kappa)$, where $T=(V,A,r)$ is a
versioned rooted Posture Tree, $\pi$ is a policy profile, and
$\kappa$ is the active control-context fingerprint. The root $r$
is the accountable claim; internal nodes represent objectives and
risks; leaves represent control tests and terminate at evidence
interfaces. Following the Composite design pattern, every node
implements the same request \cite{gamma_elements_1995}:
\[
  \mathsf{assure}(n,P,E)\rightarrow R_n .
\]
A leaf binds and verifies evidence; a Composite evaluates its
children first. $R_n$ carries state, score, child results, evidence
references, and exceptions. Tree shape and bindings are
configuration; $\pi$ supplies thresholds, weights, mandatory
branches, evidence windows, and gate consequences.

At evaluation time, $B(E,\ell,\kappa)$ admits only observations
whose scope, configuration, provenance, freshness, and integrity
match leaf $\ell$ and context $\kappa$; $B(E,\kappa)$ denotes these
bindings over all leaves. Thus
\begin{equation}
 R_r=\mathsf{ASSURE}\!\left(T,\pi,B(E,\kappa)\right).
 \label{eq:rq1_contract}
\end{equation}
Changing a model or evidence source changes $E$ or $\kappa$;
changing risk appetite or sovereignty rules changes $T$ or $\pi$;
neither changes \textsf{ASSURE}.

\paragraph{Recursive TEVV.}
Assurance Objects execute Test--Evaluate--Verify--Validate (TEVV)
by post-order traversal \cite{tabassi_ai_2023}. \emph{Test} invokes
or replays the scoped control; \emph{Evaluate} translates and
admits observations; \emph{Verify} applies the policy rule at a leaf
or branch; and \emph{Validate}, at the root, determines whether the
claim supports the intended SDLC decision. For $m_{\ell}$ admitted
observations $x_{\ell,j}\in[0,1]$, and for the admitted child set
$C_n$ with policy-normalised weights $\bar w_{n,c}$,
\begin{equation}
 s(n)=
 \begin{cases}
 \displaystyle
 \frac{1}{m_{\ell}}\sum_{j=1}^{m_{\ell}}x_{\ell,j},
 & n=\ell\text{ is a leaf},\\[2mm]
 \displaystyle
 \sum_{c\in C_n}\bar w_{n,c}s(c),
 & n\text{ is a Composite},
 \end{cases}
 \qquad
 \sum_{c\in C_n}\bar w_{n,c}=1 .
 \label{eq:rq1_recursive_score}
\end{equation}
Score and state are separate: required missing or inadmissible
evidence yields \textsc{Inconclusive}; a mandatory failure or
missed threshold yields \textsc{Fail}; an aggregate pass with a
non-mandatory exception yields \textsc{Pass with Exceptions};
otherwise it yields \textsc{Pass}. Exceptions remain attached to
the root.

\begin{center}
\begin{minipage}{0.92\linewidth}
\small\ttfamily
ASSURE(n,P,E):\\
\quad if leaf(n): return VERIFY(n, B(E,n,P.context), P.policy)\\
\quad C = [ASSURE(c,P,E) for c in children(n)]\\
\quad q = VERIFY(n, AGGREGATE(C,P.policy), C, P.policy)\\
\quad if root(n): q = VALIDATE(q,C,P.policy)\\
\quad return Result(q,C,evidence\_refs,exceptions)
\end{minipage}
\end{center}

This gives \emph{assurance from instantiation}: a control receives
$T$, $\pi$, and $\kappa$ before governing a gate and remains
\textsc{Inconclusive} until evidence binds. Context change
invalidates incompatible bindings; policy change re-evaluates
compatible evidence. Full evaluation over $M$ observations and
$N$ nodes is $O(M+N)$; for changed leaves $\Delta L$,
\begin{equation}
 T_{\mathrm{inc}}=
 O\!\left(
   \Delta M+
   \left|
     \bigcup_{\ell\in\Delta L}A(\ell)
   \right|
 \right)
 \leq O(\Delta M+kD),
 \label{eq:rq1_incremental}
\end{equation}
where $A(\ell)$ is the ancestor set, $k=|\Delta L|$, and $D$ is
tree depth. Equation~\ref{eq:rq1_incremental} gives a theoretical incremental bound. The reported RQ2 scale conditions instead force full recomputation, $O(M+N)$ per Context, with all Contexts affected.

\paragraph{Scalability comparison.} In Table \ref{tab:complexity-comparison} SMACTR incurs $O(A)$ operational effort per reassessment due to repeated artefact review, testing, and expert judgement, while AgentSentinel requires
$O(R+Q\lambda)$ to monitor runtime operations and escalate selected events to
an LLM auditor. In contrast, TAIP costs $O(M+N)$ for full evaluation and
$O(\Delta M+kD)$ for incremental updates, as bounded in
Eq.~\ref{eq:rq1_incremental}. These expressions characterise different workloads; they do not establish a measured speedup. RQ2 evaluates TAIP's full-recomputation path against a gate budget, leaving incremental performance and comparative speedup unmeasured.

\begin{table}[t]
\centering
\caption{Per-context complexity of maintaining current assurance over
RepoAudit. SMACTR denotes operational audit effort rather than CPU time.}
\label{tab:complexity-comparison}

\footnotesize
\setlength{\tabcolsep}{4pt}
\renewcommand{\arraystretch}{0.95}

\begin{tabularx}{\linewidth}{
    >{\centering\arraybackslash}m{0.17\linewidth}
    >{\raggedright\arraybackslash}X
    >{\raggedright\arraybackslash}m{0.34\linewidth}
}
\toprule
\textbf{Approach} &
\textbf{Dominant recurring work} &
\multicolumn{1}{c}{\textbf{Complexity}} \\
\midrule

SMACTR \cite{raji_closing_2020}&
Repeat artefact review, testing, and expert assessment. &
\centering $O(A)$
\tabularnewline

AgentSentinel \cite{hu_agentsentinel_2025}&
Monitor runtime operations and escalate selected events. &
\centering $O(R+Q\lambda)$
\tabularnewline

TAIP &
Bind evidence and evaluate the Posture Tree in full; selective path updates are theoretical. &
\begin{tabular}[c]{@{}l@{}}
Full (evaluated): $O(M+N)$ \\[-1pt]
Incremental (bound): \\[-1pt]
$\displaystyle
O\!\left(
\Delta M+
\left|
\bigcup_{\ell\in\Delta L}A(\ell)
\right|
\right)$ \\[-1pt]
$\displaystyle
\le O(\Delta M+kD)$
\end{tabular}
\tabularnewline

\bottomrule
\end{tabularx}

\vspace{1pt}
\begin{minipage}{\linewidth}
\scriptsize
\emph{Notation.}
$A$ is the effort of one audit engagement;
$R$ is the number of monitored operations;
$Q$ is the number of escalations;
$\lambda$ is the cost of one LLM audit;
$M$ is the number of observations;
$N$ is the number of Posture Tree nodes;
$k$ is the number of affected leaves; and
$D$ is the tree depth.
\end{minipage}
\end{table}
\subsection{TAIP Assurance Engine Implementation over RepoAudit}

RepoAudit is the unmodified operational control; the Trustworthy
AI Posture (TAIP) Assurance Engine is the assurance harness around
it. TAIP invokes pinned RepoAudit through its command-line
interface but neither parses the target repository nor selects
defects \cite{guo_repoaudit_2025}.

\begin{figure}[t]
\centering
\IfFileExists{figures/figure2-repoaudit-taip-flow.pdf}{%
  \includegraphics[width=\linewidth]{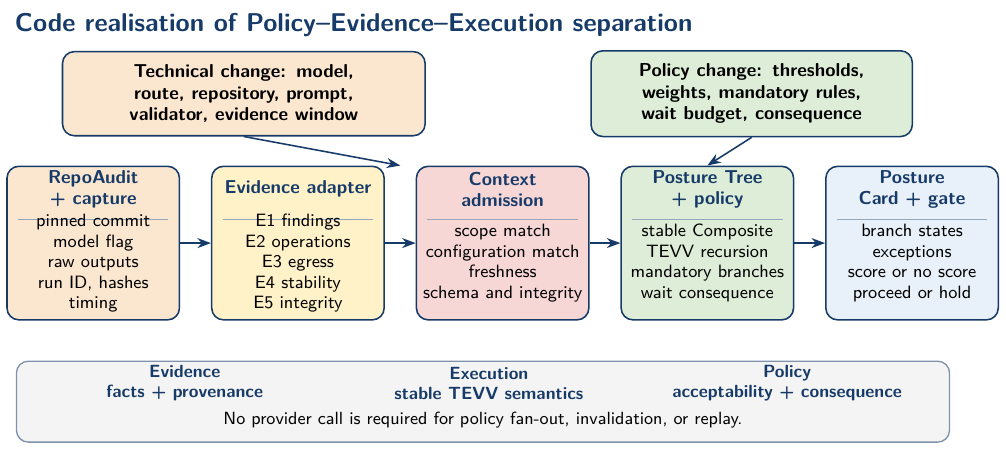}%
}{%
  \figureplaceholder{48mm}{\textbf{Figure 2 scaffold.} Pinned RepoAudit and captured raw outputs feed E1--E5 evidence translation, context admission, a versioned Posture Tree and policy profile, and a Posture Card with the gate action.}%
}
\caption{Policy--Evidence--Execution Separation realised by TAIP over RepoAudit. Evidence owns facts and provenance, execution owns stable TEVV semantics, and policy owns acceptability and consequence. Policy fan-out, invalidation, and replay require no provider call.}
\label{fig:rq1_architecture}
\end{figure}

An experiment-owned runner invokes RepoAudit in a unique directory and retains
\texttt{detect\_info.json}, \texttt{dfbscan.log}, provider-call
records, status, timing, and hashes. Its manifest records
repository, commit, bug class, model identities, provider route,
prompt, validator, and evidence window.

The \emph{Control--Test Translation and Evidence Binding Layer}
is the only component that understands RepoAudit formats.
\texttt{repoaudit\_adapter.build\_bundle} maps raw outputs into
five policy-neutral evidence families: findings, operations,
egress, stability, and integrity. \texttt{leaf\_evidence.build}
creates technical observations without encoding risk appetite.
\texttt{gate.Context} checks each observation against $\kappa$;
stale, incomplete, integrity-failed, or mismatched records remain
retained but are not admitted, producing \textsc{Inconclusive}.

The engine loads $T$ from \texttt{objectives.yaml} and $\pi$ from
\texttt{PolicyProfile}. \texttt{Node}, \texttt{ASSURE}, and
\texttt{Result} realise the Composite participants, traversal, and
exception-preserving reply. \texttt{ccpa.Controller} triggers
evaluation after evidence, execution, context, policy, or replay
events and publishes the posture card, exception path, timing, and
CI result.

Dependencies are one-way: RepoAudit knows nothing about TAIP;
the adapter knows formats but no policy; the technical oracle
establishes facts but no risk appetite; and the engine knows only
the Assurance Object contract. Format change is isolated to the
adapter, policy change to the profile, and context change to
admission. On liveness timeout TAIP publishes
\textsc{Inconclusive}; $\pi$, rather than the engine, selects a
fail-open exception or fail-closed hold.

\paragraph{Answer to RQ1.}
The pattern combines a context-bound Posture Tree, runtime
evidence binding, and Composite Assurance Objects executing
recursive TEVV. TAIP realises it around unmodified RepoAudit
through a process boundary, adapter, context gate, and stable
engine. Controls can therefore be instantiated with assurance
semantics and recomputed after evidence or policy change without
inheriting stale posture.

\section{RQ2: Can TAIP Recompute Current Posture Fast Enough for an SDLC Decision Gate?}
\label{sec:rq2}

\subsection{Decision-gate use case}
\label{sec:rq2_usecase}

This experiment considers a software development pipeline in which human developers and AI coding agents continuously submit source code to an SDLC Decision Gateway. Before each merge decision, RepoAudit operates as an AI-enabled detective security control that analyses the repository and produces security evidence. The Decision Gateway must determine whether the control can still be trusted to support the release decision under the current policy and operating context.

RepoAudit is used as the representative agentic SDLC security control, while the proposed Policy--Evidence--Execution Separation (PEES) pattern is implemented as the Trustworthy AI Posture (TAIP) assurance engine. TAIP continuously observes the evidence produced by RepoAudit, maintains the corresponding posture, and publishes a Posture Card representing the current assurance state presented to the Decision Gateway.

The operational requirement is straightforward. Whenever new evidence is produced, the governing policy changes, or the control context changes, the Decision Gateway requires an updated posture before the pipeline reaches its declared waiting limit. The experiment therefore measures a single property of the proposed assurance mechanism: the time required to recompute and publish the updated posture after a relevant change has been observed.

\textbf{RQ2.} Can TAIP recompute and publish the current posture of an AI-enabled SDLC security control within the declared waiting budget of an SDLC Decision Gateway following a policy, evidence, or observable control-context change?

The evaluation is intentionally restricted to temporal adequacy. TAIP is considered adequate when it produces a current Posture Card within the declared gate budget. RepoAudit detection accuracy, distributed deployment, and production operating cost are outside the scope of this experiment.

\begin{figure}[!t]
\centering
\IfFileExists{figures/figure3-simple-gate-experiment.pdf}{%
  \includegraphics[width=\linewidth]{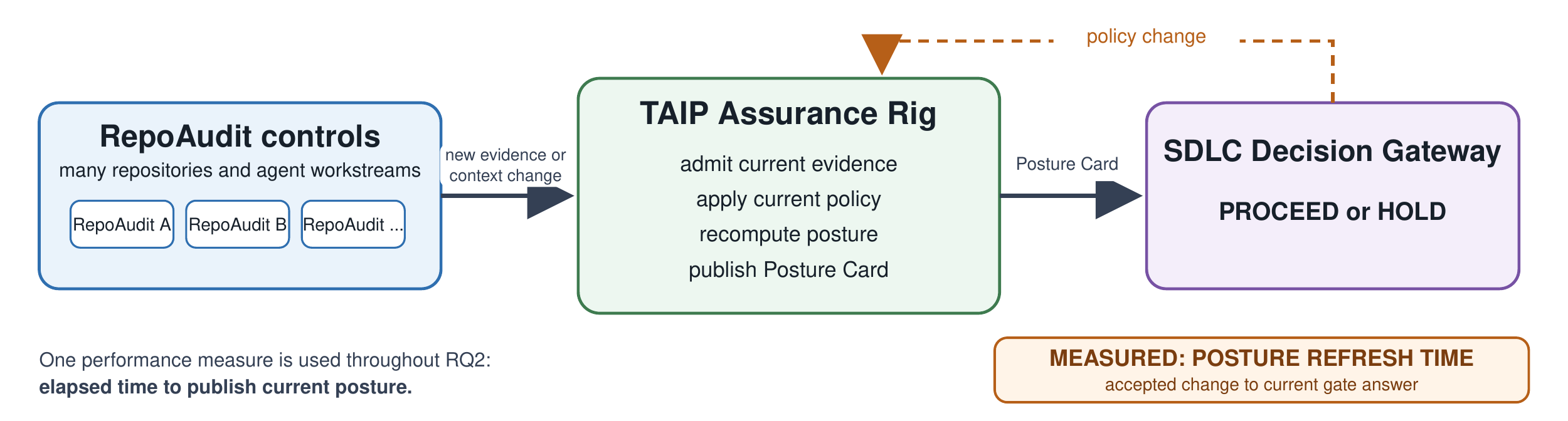}%
}{%
  \figureplaceholder{36mm}{\textbf{RQ2 experiment placeholder.} RepoAudit controls send evidence or context changes to the TAIP Assurance Rig. TAIP publishes a current Posture Card to the SDLC Decision Gateway. Policy changes can also trigger recomputation.}%
}
\caption{RQ2 experimental boundary. RepoAudit controls provide new evidence or an observable control-context change. A policy change may also originate from the Decision Gateway. TAIP recomputes posture and publishes the resulting Posture Card. The measured quantity is the elapsed time from the accepted change to the current gate answer.}
\label{fig:rq2_simple_experiment}
\end{figure}

\subsection{Experiment setup}

The experiment was executed as a single-host Python workload in an isolated sandbox\footnote{https://github.com/guylupo-stormtree/taip-repoaudit-posture}. Live provider calls were used only to acquire RepoAudit evidence; evidence translation, Context Admission, posture computation, replay, and scale measurement ran locally within the same environment. Table~\ref{tab:rq2_construction} records the execution boundary, evidence volume, and controls used to support reproducibility.

\begin{table}[!t]
\caption{Experimental execution environment and evidence volume.}
\label{tab:rq2_construction}
\centering
\scriptsize
\setlength{\tabcolsep}{4pt}
\begin{tabularx}{\linewidth}{L{0.26\linewidth}Y}
\toprule
Setup item & Experimental configuration \\
\midrule
Execution environment &
Isolated, single-host Python sandbox. The experiment did not use a distributed assurance service, external database, or separate computation cluster. \\

Runtime &
\texttt{Python 3.13.5}. RepoAudit, the evidence collector, the TAIP Assurance Engine, the invariant checks, the scale harness, and the report generator were executed as Python processes. \\

Control subject and benchmark &
RepoAudit was pinned and executed without modification at commit \texttt{160f5bc}. The audit condition was the \PrimaryConditionName{}. \\

Model coverage &
\ModelCount{} model configurations were exercised: \ModelOneName{} and \ModelTwoName{}. Each model was executed \RunsPerModel{} times. \\

Evidence volume &
The retained evidence repository contained \TotalLiveRuns{} live RepoAudit executions and \TotalLiveRuns{} run-level Evidence Bundles. Each bundle preserved findings, validator output, process telemetry, routed-model identity, timing, provider cost, manifests, checksums, and execution logs. \\

Evidence acquisition and local computation &
Model-provider calls were required for live RepoAudit evidence acquisition. Context Admission, Posture Tree evaluation, policy recomputation, model-context invalidation, replay, and Posture Card publication were performed locally. Policy-only recomputation, invalidation, and replay required no additional provider call. \\

Scale configuration &
The breadth sweep evaluated up to \MaxTestedContexts{} independent Contexts using one-worker and eight-worker configurations. The gate decision budget was declared as \GateBudget{} before measurement. \\

Reproducibility controls &
The control revision, model identity, repository scope, Policy Profile, Evidence Window, run identifier, timestamps, manifests, and hashes were retained. Each run used a separate output directory, and previous evidence was not overwritten. \\

Result provenance &
Reported timings are reconciled against \path{sequence-summary.json} and \path{scale-summary.json}. Maximum-based macros separate corrected reporting from legacy generator labels. \\
\bottomrule
\end{tabularx}
\end{table}

RepoAudit is the experimental control subject, rather than the contribution being benchmarked \cite{guo_repoaudit_2025}. The Python null-pointer-dereference benchmark provides one fixed audit condition. Repeated runs create a real Evidence Window whose findings and operational records can be admitted by TAIP. RQ2 starts after that evidence exists. RepoAudit execution time and provider latency are therefore separated from the time required to recompute posture.

\subsection{Posture-recomputation test}

The test follows one repeated sequence. RepoAudit is first executed against the NPD benchmark and its outputs are retained as Evidence Bundles. The Evidence Adapter translates those bundles into policy-neutral observations. TAIP admits observations whose scope and configuration match the active Context and computes an initial Posture Card. The harness then introduces a policy, evidence, or observable control-context change, starts the timer when TAIP accepts the event, and stops it when the replacement Posture Card is published at the local gate interface. The same operation is repeated with a full Evidence Window and with increasing numbers of independent Posture Tree instances.

Table~\ref{tab:rq2_triggers} lists the tested events. Every row exercises the same assurance operation and the same performance measure. The expected output is checked before the elapsed time is accepted. A fast result that inherits stale evidence does not pass the test.

\begin{table}[!t]
\caption{Events that trigger posture recomputation.}
\label{tab:rq2_triggers}
\centering
\small
\setlength{\tabcolsep}{4pt}
\begin{tabularx}{\linewidth}{L{0.20\linewidth}L{0.27\linewidth}Y}
\toprule
Trigger & Change introduced & Required TAIP response \\
\midrule
Policy change & The risk appetite or gate rule changes while the Evidence Window remains fixed. & Recompute the same evidence under the new Policy Profile and publish the revised Posture Card without another RepoAudit or provider call. \\
New evidence & A completed RepoAudit run adds compatible evidence to the active Evidence Window. & Admit the new evidence, recompute the Posture Tree, and publish posture based on the enlarged window. \\
Control-context change & The active model, route, repository, prompt, validator, or another fingerprinted property changes before compatible evidence is available. & Reject the inherited evidence and publish \textsc{Inconclusive}; the earlier Posture Card must not be reused. \\
Broadcast change & One policy or observable control event affects several Contexts. & Recompute each affected Posture Tree independently and publish one current Posture Card per Context. \\
\bottomrule
\end{tabularx}
\end{table}

Two workload dimensions are used. Evidence volume is increased from an initial run to the full Evidence Window of \RunsPerModel{} runs for each model. The number of Posture Tree instances is then increased to \MaxTestedContexts{}. The scale sweep uses one frozen, schema-valid evidence bundle with a distinct fingerprint for each replicated Context. This tests the cost of assurance computation. It does not represent \MaxTestedContexts{} simultaneous live RepoAudit executions or \MaxTestedContexts{} different detector workloads.

Each scale condition was repeated 30 times. The policy sequence contains three policy-class cycles (E2, E5 and E5b) in one execution. We report sample maxima and, for the eight-worker series, lower medians: the 15th ordered duration out of 30. These descriptive statistics do not establish population percentiles or a worst-case execution-time bound.

\subsection{Results}
\label{sec:results}

\begin{table}[!t]
\caption{RQ2 experimental progression and maximum observed refresh times.}
\label{tab:rq2_progression}
\centering
\small
\setlength{\tabcolsep}{4pt}
\begin{tabularx}{\linewidth}{L{0.22\linewidth}L{0.29\linewidth}L{0.20\linewidth}Y}
\toprule
Test & Evidence base & Assurance scale & Result \\
\midrule
Policy sequence & Retained compatible evidence; E2, E5, E5b & 1 Context & \PolicyLatencyMax{} (rounded); 3 cycles in one execution \\
Evidence base & \TotalLiveRuns{} executions across \ModelCount{} models & Context-bound admission & Source evidence, not timing repetitions \\
Scale policy & Frozen evidence per Context & \MaxTestedContexts{} Contexts; 1 worker & \ScaleLatencyMax{}; 30 repetitions \\
\bottomrule
\end{tabularx}
\end{table}
RQ2 evaluates whether TAIP can publish current posture within the gate budget. Retained RepoAudit evidence supported a single-context policy sequence and a repeated scale sweep. The former measures event-to-publication latency; the latter measures aggregate refresh over affected Contexts. Evidence volume and timing sample sizes are reported separately.

\paragraph{Establishing the assurance unit.}

Retained RepoAudit evidence supported the single-context policy sequence. Across its three policy-class cycles (E2, E5 and E5b), the maximum event-admission-to-publication time was 1.1309\,ms, reported as \PolicyLatencyMax{} after rounding. These cycles belong to one sequence execution. The timed path excludes repository scanning, model invocation and evidence collection, isolating the cost of recomputing and publishing a current Posture Card.

\paragraph{Expanding the evidence repository.}

A further \RunsPerModel{} RepoAudit executions were produced using
\ModelTwoName{}, increasing the retained repository to \TotalLiveRuns{}
executions across \ModelCount{} model configurations. The expanded repository
provides alternative evidence records and model-bound contexts for policy,
evidence-admission, and context-invalidation tests. Evidence acquisition is outside the timed assurance path; TAIP recomputes posture over existing, context-compatible records.

\paragraph{Aggregate scale result.}

At \MaxTestedContexts{} independent Contexts, the one-worker policy-change condition used full recomputation with all Contexts affected. Its maximum full-fleet refresh time across 30 repetitions was 1616.3\,ms, reported as \ScaleLatencyMax{}, below the \GateBudget{} budget. This is the headline scale result; it measures completion of the affected workload, not an average per-Context latency.

\begin{table}[!t]
\caption{Isolated policy-change scaling: eight workers, full recomputation, all Contexts affected; 30 repetitions per condition.}
\label{tab:rq2_isolated_scale}
\centering
\small
\setlength{\tabcolsep}{8pt}
\begin{tabular}{rrr}
\toprule
Contexts & Lower median & Maximum observed \\
\midrule
1     & 1.3\,ms   & 4.1\,ms \\
10    & 3.9\,ms   & 180.0\,ms \\
100   & 30.0\,ms  & 35.3\,ms \\
1,000 & 288.9\,ms & 317.0\,ms \\
\bottomrule
\end{tabular}
\end{table}

\begin{figure}[!t]
\centering
\includegraphics[width=\linewidth]{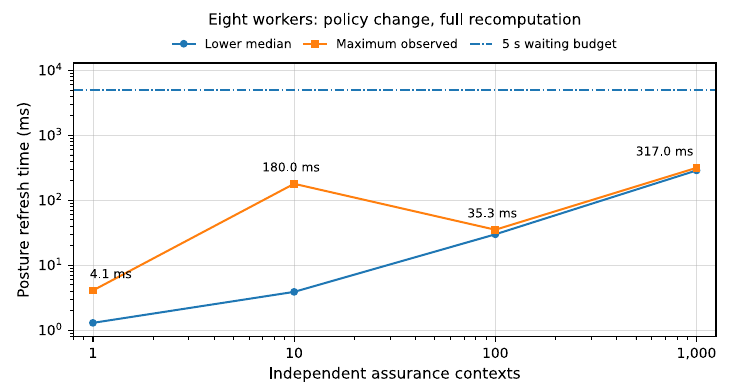}
\caption{Eight-worker policy-change scaling with full recomputation and all Contexts affected. Lower medians and observed maxima correspond to Table~\ref{tab:rq2_isolated_scale}; the one-worker headline is reported in Table~\ref{tab:rq2_progression}.}
\label{fig:rq2_isolated_scale}
\end{figure}

\paragraph{Isolated scaling behaviour.}

Table~\ref{tab:rq2_isolated_scale} and Figure~\ref{fig:rq2_isolated_scale} report the eight-worker series. Lower median refresh time increased from 1.3\,ms at one Context to 288.9\,ms at \MaxTestedContexts{} Contexts. The corresponding maxima were 4.1, 180.0, 35.3 and 317.0\,ms, all below the \GateBudget{} budget. The elevated ten-Context maximum is retained as observed; these timings alone do not identify its cause.

The 317.0\,ms maximum concerns eight workers, whereas the \ScaleLatencyMax{} headline concerns one worker. The one-Context scale cell also differs from the three-cycle policy sequence underlying \PolicyLatencyMax{}. Each value remains tied to its own workload and timing sample.

\paragraph{Current posture rather than a fast stale answer.}

Timing trials were accepted only when TAIP published the state required by
the triggering change. Applying the strict Policy Profile to unchanged
evidence moved the root posture from \BaselineRootState{} to
\StrictRootState{}. A model-context change withdrew the previously applicable
evidence and published \ModelTwoRootState{}, while replay was
\ReplayStatus{}. These checks bind the reported latency to a current
assurance decision.

\paragraph{Answer to RQ2.}

Within the tested single-host environment, the maximum observed single-context policy-to-posture latency was \PolicyLatencyMax{} (rounded). Full policy-triggered recomputation across \MaxTestedContexts{} Contexts with one worker recorded a maximum aggregate refresh of \ScaleLatencyMax{}. Both were below the \GateBudget{} budget, supporting temporal adequacy for the stated samples and execution conditions.

\paragraph{Industry implication.}

AI-enabled security controls can produce findings, reports, and alerts, but
their output does not independently establish whether the evidence remains
applicable to the current model context or governing policy. TAIP performs
that missing assurance step before the Decision Gateway consumes the result.
The measured assurance unit and aggregate scale result indicate that current,
context-bound posture can be published inside the tested pipeline waiting
budget, allowing assurance to participate in automated merge decisions rather
than remain a retrospective audit activity.

The experiment does not establish RepoAudit detection accuracy, distributed
performance, comparative speedup, or production-wide suitability. It
establishes the temporal adequacy of the assurance mechanism under the
reported single-host configuration.

\subsection{Limitations and claim boundary}

\paragraph{Sample size and interpretation.}
Three policy-class cycles in one execution and 30 repetitions per scale condition provide descriptive observations only. The reported maxima are sample extremes, not worst-case execution-time bounds or future deadline guarantees. The predeclared \GateBudget{} budget is unchanged; the original protocol's population-tail target is not established.

\paragraph{Deployment boundary.}
The experiment ran on one isolated Python host; no distributed-performance claim is made. Distributed scheduling, network delay, high availability, multi-host coordination, and production orchestration remain future work.

\paragraph{Replicated Contexts.}
They are replicated from one frozen, schema-valid Evidence Bundle and assigned distinct fingerprints. The claim concerns TAIP assurance computation over independent Posture Tree instances. It does not concern detector diversity or \MaxTestedContexts{} concurrent live RepoAudit runs.

\paragraph{Comparison boundary.}
There is no manual, passive-monitoring, or competing-posture-engine baseline. This is deliberate. The experiment reports the absolute time required by the proposed mechanism against a gate budget, rather than a speedup over an alternative.

The result is also bounded to one Python NPD condition, one pinned RepoAudit revision, two model configurations, one Evidence Window design, and one local execution environment. It evaluates posture recomputation, not universal RepoAudit accuracy, complete defect detection, production CI availability, developer waiting experience, or the organisational effect of repeated holds. TAIP can reject evidence that is stale, incomplete, or bound to another Context, but it cannot correct an erroneous benchmark label, an incorrect adapter, or an unsupported source finding.

\section{Conclusion}
\label{sec:conclusion}

This paper introduced the Policy--Evidence--Execution Separation pattern and its TAIP Assurance Engine implementation for continuous, context-bound assurance of agentic security controls at software-delivery gates. Separating stable assurance execution from versioned policy and admissible evidence allows posture to be recomputed without re-running the underlying auditor. Using retained RepoAudit evidence, the maximum observed single-context policy-to-posture latency was \PolicyLatencyMax{} (rounded). Full policy-triggered recomputation across \MaxTestedContexts{} independent assurance Contexts with one worker recorded a maximum aggregate refresh of \ScaleLatencyMax{}, below the predeclared \GateBudget{} Decision Gateway budget. These observations support inline assurance within the evaluated single-host conditions. The contribution is an assurance architecture that checks whether evidence remains current and context-compatible before a software-delivery decision.

\subsection{Future directions}

A potential future direction is to evaluate our approach across different agentic security controls, repository types, model providers, and security tasks. Such studies should assess not only recomputation latency, but also the correctness of evidence admission, resistance to malformed or adversarial evidence, and the operational effects of false holds, exceptions, and inconclusive decisions on development workflows.

Future work includes incremental, dependency-aware recomputation. Implementations could identify shared evidence and policy dependencies, update only affected Posture Tree paths, and measure whether this reduces latency and memory use at enterprise scale.

\begin{credits}

\subsubsection{Generative AI Use Disclosure.}
Generative AI tools assisted language revision, architectural
reorganisation, and code-planning review. The authors remain
responsible for the research design, code, source verification,
experimental execution, retained data, statistical analysis,
and all claims in the manuscript.

\subsubsection{\discintname}
The authors have no competing interests to declare that are
relevant to the content of this article.

\end{credits}

\bibliographystyle{splncs04}
\bibliography{reference}

\end{document}